\documentclass[fleqn,twoside]{article}
\usepackage{espcrc2}
\usepackage{epsfig}
\usepackage{mathbbol}

\newcommand{\eq}{\begin{equation}}
\newcommand{\en}{\end{equation}}
\newcommand{\eqa}{\begin{eqnarray}}
\newcommand{\ena}{\end{eqnarray}}
\newcommand{\be}{\begin{equation}}
\newcommand{\ee}{\end{equation}}
\newcommand{\ba}{\begin{eqnarray}}
\newcommand{\ea}{\end{eqnarray}}

\newcommand{\ZZ}{\hbox{{\rm Z{\hbox to 3pt{\hss\rm Z}}}}}

\newcommand{\AmS}{{\protect\the\textfont2
  A\kern-.1667em\lower.5ex\hbox{M}\kern-.125emS}}

\title{
Speeding up the Hybrid-Monte-Carlo algorithm for dynamical fermions}
\author{M. Hasenbusch 
\thanks{presented by M. Hasenbusch} and
K. Jansen 
\phantom{xxxxxxxxxxxxxxxxxxxxxxxxxxxxxxxxxxxxxxxxxxxxxxxx} 
{\small DESY 01-155}\\
NIC/DESY Zeuthen,
Platanenallee 6,  D-15735 Zeuthen, Germany }
\begin{document}
\begin{abstract}
We propose a modification of the Hybrid-Monte-Carlo algorithm that allows for
a larger step-size of the integration scheme at constant acceptance rate.
The key ingredient is the splitting 
of the pseudo-fermion action into two parts. We test our proposal
at the example of the two-dimensional lattice Schwinger model and four-dimensional
lattice QCD
with two degenerate flavours of Wilson-fermions.
\end{abstract}
\maketitle
\section{INTRODUCTION}
In large scale simulations of lattice QCD with two flavours of mass-degenerate
Wilson fermions, the mass of the fermions is still too large 
compared with the up and the down  quark masses. Therefore a delicate 
extrapolation  of the data is needed.
Unfortunately, the numerical effort required for the Hybrid-Monte-Carlo (HMC)
\cite{hybrid} algorithm increases as the quark mass
decreases for, at least, three reasons \cite{QCDwith,Jansenrev}:
the solver (BiCGstab, conjugate gradient) needs more iterations,
the step-size of the integration scheme has to be
reduced to maintain a given acceptance rate, the autocorrelation times 
(in units of trajectories) increase.


%
We propose to modify the pseudo-fermion action in such a way that the 
problem of the decreasing step-size is drastically reduced.
Our starting point is the observation that the step-size can be increased when
the fermion matrix is replaced by its preconditioned counter-part 
\cite{Pe,alphabench}. This means that reducing the condition number of the 
fermion matrix allows for a larger step-size. 

Here we shall present results
for the two-dimensional Schwinger model \cite{MH_schwinger} and 
four-dimensional lattice QCD.

\section{THE SCHWINGER MODEL}
Let us start with the partition function 
with two degenerate 
flavours of Wilson-fermions
\be
Z = \int \mbox{D}[U] \exp[-S_G(U)] \; \; \mbox{det} M(U)^\dag M(U) ,
\label{action}
\ee
where $S_G(U)$ is the gauge action and $M(U)=\mathbb{1}-\kappa H(U)$
the fermion matrix.
In the HMC, the fermion determinant is 
represented by an integral over a so-called pseudo-fermion field $\phi$:
\be
\mbox{det} M^\dag M \; \propto \; \int \mbox{D}[\phi] \mbox{D}[\phi^\dag] \;
\exp(-|M^{-1} \phi|^2) .
\ee
Hence the pseudo-fermion action is given by $S_F=|M^{-1} \phi|^2$.
The key-ingredient of our method is to split $M$ into two factors, where each
factor has a reduced condition number. For each factor we use a pseudo-fermion
field:
\begin{equation}
S_{F1} = |\tilde M^{-1} \psi|^2  \;\;\;\;\;\;\; 
S_{F2} = |\tilde M M^{-1} \phi|^2  \;\;,
\end{equation}
where $\tilde M = \mathbb{1} - \tilde \kappa H$
with $0 \le \tilde \kappa \le \kappa$. 

In the HMC we have to compute the variation of the action with respect to the 
gauge-field. It turns out that for our modified action this can be done 
in much the same way as for the standard pseudo-fermion action:

\begin{equation}
\delta S_{F2}(U,\phi) =
-\left[Y^\dag \;\delta M \;X + X^\dag \; \delta M^\dag \; Y \right] \;\;
\end{equation}
with
$X=M^{-1} \phi$ and $Y=M^{\dag -1} (a \phi + b X)$,
where $a = \tilde \kappa/\kappa$  and $b = 1-a$.

In our numerical study of the Schwinger model,
we have applied this modification on top of even-odd
preconditioning:
$M$ is replaced by
 $M_{ee} = \mathbb{1}_{ee}-\kappa^2 H_{eo} H_{oe}$ .
The modified pseudo-fermion action is given by
\begin{equation}
 S_{F1} = |\tilde M_{ee}^{-1} \psi|^2 \;\;\;\; \;\;\;\;\;
 S_{F2} = |\tilde M_{ee} M_{ee}^{-1} \phi|^2 \;\;,
\end{equation}
where
$ \tilde M_{ee} = \mathbb{1}_{ee}-\tilde \kappa^2 H_{eo} H_{oe}$ .
We have performed our simulations of the Schwinger model 
at the same parameters as ref. \cite{Pe}. We have used the 
leap-frog integration-scheme. In order
to separate off effects from the gauge action,
we used a reduced step-size $(n=4)$
for the gauge action \cite{SeWe}. The length of the trajectory is
taken randomly between 0.5 and 1.5. We used the BiCGstab solver to compute
$\tilde M_{ee}^{-1} \psi$ and $M_{ee}^{-1} \phi$.

\begin{figure}
\begin{center}
\includegraphics[width=8cm]{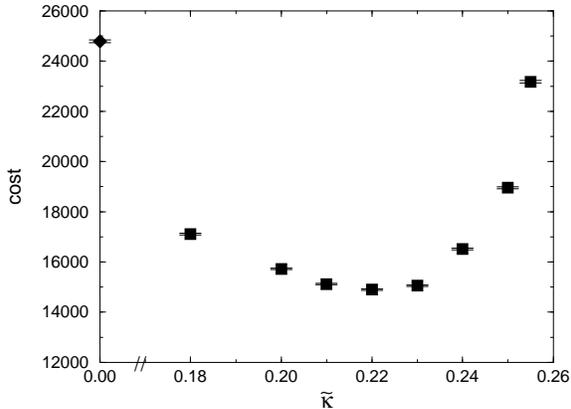}
\caption[Binder Cumulant $U$ at $Z_a/Z_p$ fixed 1]
{\label{costs} \small \\
cost = (number of applications of $H_{eo}H_{oe}$ per trajectory)
as a function of $\tilde \kappa$
for $L=32$, $\beta=4.0$ and $\kappa=0.26$. 
The
acceptance rate is fixed to $\approx 0.8$.
}
\end{center}
\end{figure}

In all cases we have chosen $\beta=4.0$. In a first set of runs we 
simulated a $32 \times 32$ lattice at $\kappa=0.26$ 
and various values of $\tilde \kappa$. 
Following ref. \cite{Pe}
the pseudo-meson mass is $m_P = 0.210(3)$ for this value of $\kappa$.
Each run consists of 
10000 trajectories. 1000 trajectories were discarded for equilibration.
We have measured the value of square Wilson-loops up to the size $5 \times 5$ 
and the topological charge with the geometrical definition.
We found that for fixed acceptance rate,
the integrated auto-correlation times of these
quantities do not depend on the value of $\tilde \kappa$, within error-bars. 
Therefore, the performance of the algorithm can be measured just by the number 
of applications of $H_{eo} H_{oe}$ that is needed in average for one 
trajectory. This number is given as  a function of $\tilde \kappa$ in figure 
\ref{costs}. We see that the numerical cost has a shallow minimum at 
$\tilde \kappa \approx 0.22$.  At this minimum cost is reduced by 
a factor of about $1.66$ compared with the standard HMC 
($\tilde \kappa=0$).

Next we performed simulations of a $64 \times 64$ lattice at 
$\kappa=0.2570$ and 
$\kappa=0.2605$. The  pseudo-meson masses at these values of $\kappa$ are 
$m_P = 0.210(3)$ and $m_P = 0.124(5)$, respectively  \cite{Pe}.
For $\kappa=0.2605$ we found a reduction of the cost by factor of $2$ compared
with the standard pseudo-fermion action. 
For $\kappa=0.257$ we see only a tiny improvement. This observation indicates 
that the gain in performance increases as we approach $\kappa_c$. 
For details see ref. \cite{MH_schwinger}.

\section{LATTICE QCD}
We have performed some exploratory runs for QCD in four dimensions. 
Here we used a different set-up as in the case of the Schwinger model. 
These changes are mostly motivated by the existing HMC-code of the 
ALPHA-collaboration.

We considered the Hermitian fermion matrix \\
$
Q = c_0 \gamma_5 M
$ .
We define 
$\tilde Q  = Q + i \sigma \; \mathbb{1}$.
That means that the modified pseudo-fermion action consists of the two parts
\begin{equation}
\label{QCDsplit1}
S_{F1} = \psi^{\dag} [Q^2 + \sigma^2  \mathbb{1}]^{-1} \psi ,
\end{equation}
\begin{equation}
\label{QCDsplit2}
S_{F2} = \phi^{\dag} [\mathbb{1} + \sigma^2 (Q^2)^{-1}]  \phi .
\end{equation}


In our simulations we have used $O(a)$ clover improvement and 
even-odd preconditioning \cite{JansenLiu}. 

Instead of the leap-frog scheme we used a scheme proposed in ref. 
\cite{SeWe} that partially removes the $\Delta \tau^2$ errors.
This integration scheme is characterised by the coefficients
1/6,  1/2,   2/3,   1/2 and  1/6. Note that in contrast to the leap-frog 
scheme, 
the derivative  of the fermion action with respect to the gauge-field 
has to be computed twice per elementary step.

\noindent
In a first set of runs we studied an  $L^4$ 
system with Schr\"odinger-functional (SF)
boundary conditions,
where $\kappa \approx \kappa_c$.  The parameter $\sigma$ of the modified 
action is set as $\sigma^2 = \sqrt{\lambda_{min}}$,
where $\lambda_{min}$ is the smallest eigenvalue of $\hat Q^2$. 
Our results are 
summarised in table \ref{SFtable}. 
For $L=8$ the step-size is enlarged by a factor
of $1.4$ by the modification of the action, 
while the overhead is about $1.29$.
For $L=12$, we  see already an increase of the step size by a factor of $2$,
while the over-head is $1.28$.

\begin{table}
\caption{\sl \label{SFtable} 
Runs with SF-boundary conditions.
$\Delta \tau$ is the step-size of the integration-scheme, $N_{md}$ the number 
of steps per trajectory, ``acc" the acceptance rate. 
$N_{CG}$ and $N_{CG2}$ are the average number of iterations needed to compute 
$(\hat Q^2)^{-1} \phi$ and $[\hat Q^2 + \sigma^2  \mathbb{1}]^{-1} \psi$
with the conjugate gradient solver.
The standard 
HMC data are taken from ref. \cite{alphabench}.
}
\begin{tabular}{rllrlrr}
\hline
$L$ &$\beta$ & $\Delta \tau$ & $N_{md}$ & acc  & $N_{CG}$ & $N_{CG2}$ \\
\hline
8 & 7.2103 & 0.14     &  7      & $92 \%$ &   79   &   23  \\
8 & 7.2103 & 0.10     & 10      & $92 \%$ &   80   &       \\
\hline
12  &7.5 & 0.14        &  7     & $84 \%$ &  114   &   32  \\
12  &7.5 & 0.075       & 13     & $89 \%$ &  114   &       \\
\hline
\end{tabular}
\end{table}

Finally we performed runs at a smaller value of $\beta$ for a physically 
large volume. We took the parameters for the run from ref. \cite{Sroczynski}.
In ref. \cite{Sroczynski} the leap-frog scheme was used. 
For  $\Delta \tau= 0.02$
and $N_{md}=50$ elementary steps per trajectory they obtained an acceptance 
rate of $80 \%$. 

Our results are given in table  \ref{Largetable}.
For these particular parameters, the improved integration scheme performs better
than the leap-frog. With the standard pseudo-fermion action we get an 
acceptance rate of $82 \%$  with $\Delta \tau = 0.06 $. I.e.  the improved
scheme out-performs the leap-frog by a factor of 1.5. Using the modified 
action, the step-size can be further increased to $0.1$, maintaining an 
acceptance rate of $80 \%$.

\begin{table}
\caption{\sl \label{Largetable}
Runs for $\beta=5.2$, $\kappa=0.1370$ and $c_{sw}=1.76$ with
periodic space and anti-periodic time boundary conditions on a $8^3 \times 24$
lattice.
}

\begin{center}
\begin{tabular}{lllll}
\hline
 $\Delta \tau$ & $N_{md}$ & acc  & $N_{CG}$ & $N_{CG2}$ \\
\hline
 0.06     & 16      & $82 \%$ &   122  &       \\
 0.1      & 10      & $80 \%$ &   122  &   27    \\
\hline
\end{tabular}
\end{center}
\end{table}
\section{CONCLUSIONS}
We propose to use a modified pseudo-fermion action in the 
HMC simulation of dynamical Wilson-fermions.
It is easy to incooperate this modification in an existing 
HMC Code.  
We have demonstrated that the modification reduces the numerical costs
for the 2D Schwinger model as well as 4D QCD with
clover-improvement.
We see an improvement up to a factor of 2. However, we expect that for 
lighter quark masses this gain becomes even larger.

There  remain a number of open problems. We need longer runs for QCD to
determine reliable autocorrelation times. We have to compare the modification 
of the pseudo-fermion action with $\tilde M$ and $\tilde Q$ in the same model.
Is it useful to split the pseudo-fermion action in more than two parts?
We have to study more carefully the inter-play of various integration
schemes with the modification of the action.

Finally, our idea can be combined with Polynomial-Hybrid-Monte-Carlo algorithm
\cite{PHMC1,PHMC2}.
(See Mike Peardons plenary talk).
The implementation is straight forward. The non-trivial question is the choice
of the polynomials.
\section{ACKNOWLEDGEMENT}
We like to thank R. Sommer for discussions.

\end{document}